\documentclass[conference]{IEEEtran}
\IEEEoverridecommandlockouts
\usepackage{cite}

\usepackage{amsmath,amssymb,amsfonts}
\usepackage{algorithmic}
\usepackage{graphicx}
\usepackage{textcomp}
\usepackage{xcolor}
\usepackage{url}
\usepackage{listings}
\usepackage{soul}
\usepackage{booktabs}
\usepackage{graphicx}
\usepackage{pgfplots}
\usepackage{siunitx}
\usepackage{float}

\pgfplotsset{compat=1.18}

\newcommand{\changed}[1]{\st{#1}}

\renewcommand{\changed}[1]{}

\def\BibTeX{{\rm B\kern-.05em{\sc i\kern-.025em b}\kern-.08em
    T\kern-.1667em\lower.7ex\hbox{E}\kern-.125emX}}
\begin{document}

\title{Quantifying the Relationship Between Team Dysfunctions and Performance in Capstone Projects}

\author{\IEEEauthorblockN{Luciano Pereira Soares}
\IEEEauthorblockA{\textit{Engineering and Computer Science} \\
\textit{Insper}\\
São Paulo, Brazil \\
lpsoares@insper.edu.br}
\and
\IEEEauthorblockN{Luciana Campos Lima}
\IEEEauthorblockA{\textit{Bussines and Economics} \\
\textit{Insper}\\
São Paulo, Brazil \\
lucianacl@insper.edu.br}
\and
\IEEEauthorblockN{Rafael Corsi Ferrão}
\IEEEauthorblockA{\textit{Engineering and Computer Science} \\
\textit{Insper}\\
São Paulo, Brazil \\
rafael.corsi@insper.edu.br}
}

\maketitle

\begin{abstract}
This paper investigates the association between team dysfunctions and performance in the context of an engineering Capstone program. Capstone programs are considered fundamental in assessing the readiness of engineering and computer science students for professional practice, evaluating not only technical knowledge but also essential competencies such as teamwork, communication, design, and organization. Despite the recognized importance of team dynamics in such settings, their relationship with project performance remains insufficiently understood. Using survey data collected from student teams, we quantify team dynamics based on Lencioni’s model of the five dysfunctions of a team, with all measures reverse-coded so that higher values indicate more positive team behaviors (i.e., lower dysfunction). The results indicate a weak-to-moderate relationship between team dynamics and performance. Each dysfunction score was computed as the sum of three corresponding survey items. All items were coded such that higher values indicate more positive team behaviors (i.e., lower dysfunction). Therefore, each score ranges from 3 to 9, where higher values reflect healthier team dynamics. This study analyzes empirical evidence on team effectiveness in project-based learning environments and offers insights for the design of Capstone programs. In particular, it highlights the importance of goal alignment and focus on deliverables while emphasizing that team dynamics should be considered alongside technical and contextual factors.
\end{abstract}


\section{Introduction}
\label{sec:introduction}

Capstone programs have become an important component of engineering and computer science education, particularly in institutions that emphasize strong integration between academic training and professional practice~\cite{whalley_computing_2025,KhakurelPorras2020}. These programs aim to bridge the gap between theoretical knowledge and real-world application by engaging students in complex, open-ended projects, often in collaboration with industry partners~\cite{jamaluddin_undergraduate_2013,nasir_software_2025}. In this context, students are not only required to demonstrate technical proficiency but also to develop essential professional competencies such as teamwork, communication, organization, and problem-solving.

This study was conducted during the 2025 academic year in the Engineering (Computer, Mechatronics, and Mechanical) and Computer Science Capstone programs at Insper~\cite{soares2018}. These are full-time programs with a strong focus on professional preparation, emphasizing both technical excellence and the continuous development of transversal skills. The Capstone experience, which takes place in the penultimate year, plays a key role in this process. Over the course of a semester, student teams work closely with partner organizations to develop technological solutions to real and relevant problems. These projects provide students with practical exposure to industry environments, including interactions with technical mentors and academic advisors, while also enabling organizations to explore innovative ideas, prototype solutions, and address concrete challenges.

Within this setting, teamwork emerges as a critical determinant of project success~\cite{cecilia_bastarrica_impact_2023, FriessGoupee2020}. Capstone projects are typically carried out in small groups (three to four students in this study), where coordination, shared responsibility, and collective decision-making are essential. However, while technical outcomes are formally evaluated through multiple dimensions, such as execution, design, and entrepreneurship~\cite{lopez_improving_2025}, the internal dynamics of teams are often less systematically studied. In particular, it remains an open question to what extent team effectiveness, as perceived by students themselves, is associated with measurable project performance.

A widely adopted framework for understanding team effectiveness is Lencioni’s model of the Five Dysfunctions of a Team~\cite{lencioni2002five}, which identifies absence of trust, fear of conflict, lack of commitment, avoidance of accountability, and inattention to results as key barriers to team performance. Although this model is extensively used in professional and educational settings, empirical evidence of its predictive power in academic project-based environments remains limited~\cite{garcia_application_2024}.

This study investigates the relationship between perceived team dysfunctions and project performance in the context of Insper’s Capstone program~\cite{ferrao2025}. 

Using survey data collected from students during mentorship activities, and aggregated at the group level, we quantify the presence of each dysfunction and analyze its correlation with project grades across multiple evaluation dimensions. By doing so, we aim to assess whether Lencioni’s framework provides explanatory power in this educational context and to identify which aspects of team dynamics are most strongly associated with successful project outcomes.

The results contribute to both educational practice and research by providing empirical insights into the role of team dynamics in project-based learning environments. In particular, they offer guidance for instructors and program designers seeking to improve team formation, mentoring strategies, and evaluation methods in Capstone courses.

\section{Related Work}

Prior work in engineering and computing education has explored team performance, peer evaluation, and group dynamics in project-based learning environments, particularly in Capstone courses. These courses are widely used to develop both technical and professional skills through collaboration on real-world projects~\cite{whalley_computing_2025,KhakurelPorras2020}.

Team performance has been linked to factors such as communication, coordination, and teamwork quality~\cite{cecilia_bastarrica_impact_2023}. However, empirical results are often mixed. For example, some studies report moderate correlations between team characteristics and outcomes, while others find limited or non-significant relationships~\cite{grasl_exposing_2023}. Similarly, peer evaluation is commonly used to assess teamwork and promote accountability, but it presents challenges in fairness and reliability~\cite{FriessGoupee2020,porquet-lupine_evaluating_2023}.

Capstone experiences are often designed not only to develop technical skills but also to support teamwork through structured activities and mentorship, which require careful pedagogical design~\cite{paretti13}. 

A widely adopted framework for understanding team effectiveness is Lencioni’s model of the Five Dysfunctions of a Team~\cite{lencioni2002five}, which includes absence of trust, fear of conflict, lack of commitment, avoidance of accountability, and inattention to results. The model emphasizes trust as the foundation of effective teamwork, while the remaining dimensions describe how teams coordinate, make decisions, and focus on outcomes. 

Prior work~\cite{6462328} has identified common team dysfunctions based on student perceptions; however, limited evidence exists regarding their relationship with measurable performance outcomes. Applications of Lencioni’s model in practice are qualitative and case-based, with limited empirical validation in educational settings~\cite{edwards2012team}. 

Despite its widespread use, empirical validation of Lencioni’s model in educational settings remains limited. A recent mapping study highlights the lack of strong empirical evidence linking collaborative learning constructs to measurable performance outcomes~\cite{garcia_application_2024}.

This study addresses this gap by providing a quantitative analysis of the relationship between team dysfunctions and performance in a Capstone program.

\section{Methodology}

This study investigates the relationship between team dysfunctions and project performance in the context of a Capstone program. The methodology consists of three main stages: data collection, construction of dysfunction metrics, and statistical analysis.

\subsection{Data Collection}

Data were collected from student teams participating in the Capstone program. Each team consisted of three to four students working collaboratively on a project proposed by an external organization. Approximately midway through the semester, students were asked to complete a survey before participating in a teamwork mentorship session. The collected data were subsequently used to support feedback during these sessions.

The survey included 15 questions related to team behavior, each answered using a three-level Likert scale: \textit{Low}, \textit{Medium}, and \textit{High}. The questions were adapted and translated to reflect key aspects of team interaction, such as communication, accountability, and collaboration.

In total, data from 40 teams were collected from a  single cohort in spring-2025. Individual responses were aggregated at the team level to enable comparison with group performance. 

\subsection{Mapping to Team Dysfunctions}

The Capstone provides three mentoring meetings, the data was collected in one of them, in a thirty-minute meeting each group to support the reflection about the group dynamic, totaling sixteen hours. The survey, presented to the respondents as a “team assessment”, identified the Lencioni’s five dys-functions of a team: (i) absence of trust, (ii) fear of conflict, (iii) lack of commitment, (iv) avoidance of accountability, and (v) inattention to results. Each qualitative response was converted into a numerical value: Low = 1; Medium = 2; High = 3.

Each dysfunction score was computed as the sum of three corresponding survey items. Therefore, each dysfunction score ranges from 3 to 9, where higher values indicate more positive team behaviors (i.e., lower dysfunction). Scores between 3 and 5 indicate a serious problem in that dimension, scores between 6 and 7 suggest a potential problem, and scores between 8 and 9 indicate that the dysfunction is unlikely to be a concern.

Group-level scores were obtained by averaging individual dysfunction scores across all members of each team.

\subsection{Performance Measures}

Team performance was evaluated using six competency dimensions assessed at the group level.

\begin{itemize}
    \item Technical Execution
    \item Organization
    \item Communication
    \item Teamwork
    \item Design
    \item Entrepreneurship
\end{itemize}

Each dimension was graded using a letter scale (e.g., A, B+, C), which was converted into a numerical score in the range from 0 to 10 using a predefined mapping from letter grades to numeric values. We are only using the final grade for each of the elements.

An overall performance score for each team was computed as the average of the six competency scores, providing a balanced measure of both technical and professional performance.

\subsection{Statistical Analysis}

The relationship between team dysfunctions and performance was analyzed using correlation analysis. Pearson correlation coefficients were computed between each dysfunction score and the overall team performance. Spearman rank correlations were also calculated to account for potential non-linear relationships.

In addition, an overall team-health index was computed as the average of the five dysfunction scores, providing a single measure of team effectiveness. This index was also correlated with performance.

The analysis focuses on identifying which dimensions of team dynamics are most strongly associated with project outcomes, as well as assessing the extent to which Lencioni’s model explains performance variability in a project-based learning environment.

\section{Results}

A total of 40 groups had complete information for both the team-dynamics survey and the project grades; the analysis below is therefore based on these 40 matched cases. The questionnaire items are positively phrased, so higher values indicate more favorable team behavior and, in practice, lower dysfunction. For reporting purposes, we aggregated the five team-dynamics dimensions into a team-health index computed as the simple mean of the five dimension scores. Project performance was summarized by the average of the six graded competencies (technical execution, organization, communication, teamwork, design, and entrepreneurship).

\subsection{Descriptive statistics}

\begin{table}[t]
\centering
\caption{Descriptive statistics for the team-dynamics dimensions (n = 40).}
\label{tab:team_stats}
\begin{tabular}{lrrrr}
\toprule
Dimension & Mean & SD & Min & Max \\
\midrule
Absence of Trust & 7.42 & 0.91 & 5.00 & 9.00 \\
Fear of Conflict & 8.36 & 0.76 & 6.00 & 9.00 \\
Lack of Commitment & 8.14 & 0.88 & 6.00 & 9.00 \\
Avoidance of Accountability & 7.17 & 0.98 & 4.50 & 9.00 \\
Inattention to Results & 6.72 & 0.84 & 5.00 & 8.50 \\
Team-health index & 7.56 & 0.61 & 5.60 & 8.70 \\
\bottomrule
\end{tabular}
\end{table}

\begin{table}[t]
\centering
\caption{Descriptive statistics for the project grades (n = 40).}
\label{tab:grade_stats}
\begin{tabular}{lrrrr}
\toprule
Competency & Mean & SD & Min & Max \\
\midrule
Technical & 8.17 & 1.45 & 5.00 & 10.00 \\
Organization & 7.82 & 1.60 & 3.67 & 9.80 \\
Communication & 8.12 & 1.52 & 3.00 & 10.00 \\
Teamwork & 8.07 & 1.89 & 3.00 & 10.00 \\
Design & 8.01 & 1.34 & 4.67 & 10.00 \\
Entrepreneurship & 7.89 & 1.28 & 5.00 & 9.83 \\
Average grade & 8.02 & 1.31 & 4.67 & 9.80 \\
\bottomrule
\end{tabular}
\end{table}

The team dynamics scores were generally high (Tab. \ref{tab:team_stats}), with \textit{Fear of Conflict} ($M = 8.36$, $SD = 0.76$) and \textit{Lack of Commitment} ($M = 8.14$, $SD = 0.88$) showing the highest averages. \textit{Inattention to Results} was the lowest-scoring dimension ($M = 6.72$, $SD = 0.84$), while \textit{Avoidance of Accountability} also showed relatively greater dispersion ($SD = 0.98$). On the performance side, the average grade across the six competencies was 8.02 ($SD = 1.31$) as detailed in Tab. \ref{tab:grade_stats}, with \textit{Teamwork} receiving the highest mean score ($M = 8.08$, $SD = 1.89$) and \textit{Organization} the lowest ($M = 7.82$, $SD = 1.60$).

\subsection{Association between team dynamics and overall performance}

\begin{table}[t]
\centering
\caption{Correlation between each team-dynamics dimension and the average project grade.}
\label{tab:corr_avg}
\footnotesize
\setlength{\tabcolsep}{3pt}
\begin{tabular}{lrrrr}
\toprule
Dimension & Pearson $r$ & $p$-value & Spearman $\rho$ & $p$-value \\
\midrule
Absence of Trust & 0.169 & 0.297 & 0.244 & 0.130 \\
Fear of Conflict & 0.071 & 0.664 & 0.079 & 0.628 \\
Lack of Commitment & 0.207 & 0.199 & 0.171 & 0.291 \\
Avoidance of Accountability & 0.185 & 0.252 & 0.231 & 0.151 \\
Inattention to Results & 0.323 & 0.042 & 0.271 & 0.090 \\
Team-health index & 0.276 & 0.084 & 0.297 & 0.063 \\
\bottomrule
\end{tabular}
\end{table}

At the aggregate level (Tab. \ref{tab:corr_avg}), the strongest association with performance was observed for the \textit{Inattention to Results} dimension ($r = 0.323$, $p = 0.042$), followed by the overall \textit{team-health index} ($r = 0.276$, $p = 0.084$). The other dimensions showed positive but weaker and non-significant associations with the average grade. These results suggest that, in this Capstone context, the team-dynamics survey is modestly associated with performance, but the effect is modest rather than deterministic.

\subsection{Relationship between team dynamics and each graded competency}

\begin{table*}[t]
\centering
\caption{Pearson correlations between team-dynamics dimensions and the six graded competencies plus the overall average grade.}
\label{tab:corr_matrix}
\resizebox{\textwidth}{!}{%
\begin{tabular}{lrrrrrrr}
\toprule
& Technical & Organization & Communication & Teamwork & Design & Entrepreneurship & Average \\
& execution & & & & & & grade \\
\midrule
Absence of Trust & 0.024 & 0.065 & 0.273 & 0.326 & 0.036 & 0.088 & 0.169 \\
Fear of Conflict & -0.024 & 0.051 & 0.203 & 0.144 & 0.001 & -0.056 & 0.071 \\
Lack of Commitment & 0.111 & 0.081 & 0.307 & 0.368 & 0.054 & 0.084 & 0.207 \\
Avoidance of Accountability & -0.049 & 0.125 & 0.369 & 0.423 & -0.041 & 0.018 & 0.185 \\
Inattention to Results & 0.275 & 0.260 & 0.291 & 0.322 & 0.235 & 0.282 & 0.323 \\
\bottomrule
\end{tabular}}
\end{table*}

The competency-level matrix (Tab. \ref{tab:corr_matrix}) shows that the team-dynamics dimensions relate most clearly to the Teamwork grade, where Absence of Trust ($r = 0.326$), Lack of Commitment ($r = 0.368$), Avoidance of Accountability ($r = 0.423$), and Inattention to Results ($r = 0.322$) all display positive associations. Communication also exhibits a moderate relation with Lack of Commitment ($r = 0.307$) and Avoidance of Accountability ($r = 0.369$). In contrast, Technical, Design, and Entrepreneurship show weaker links to the team-dynamics measures. This pattern indicates that the survey primarily captures interpersonal and coordination aspects of the project experience rather than purely technical performance.

\subsection{Overall performance}

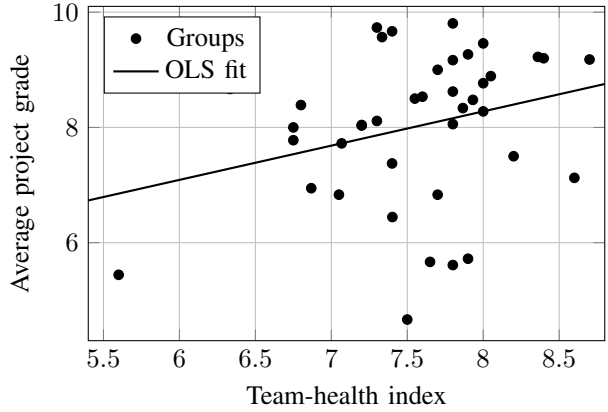
\begin{figure}[!ht]
\centering
\begin{tikzpicture}
\begin{axis}[
    width=0.95\linewidth,
    height=6cm,
    grid=both,
    xlabel={Team-health index},
    ylabel={Average project grade},
    xmin=5.4, xmax=8.8,
    ymin=4.3, ymax=10.1,
    legend style={at={(0.03,0.97)},anchor=north west},
]
\addplot[only marks, mark=*, mark size=1.8pt] coordinates {
(7.402,6.445)
(6.750,7.778)
(7.800,9.805)
(8.200,7.500)
(7.600,8.533)
(7.200,8.042)
(7.650,5.667)
(7.200,8.033)
(7.900,5.722)
(7.800,8.057)
(7.500,4.667)
(5.600,5.443)
(8.398,9.200)
(7.334,9.567)
(7.400,7.375)
(7.800,5.612)
(7.800,8.622)
(8.600,7.125)
(8.000,8.278)
(7.932,8.478)
(7.866,8.335)
(7.550,8.500)
(8.000,8.767)
(6.800,8.388)
(8.700,9.178)
(7.800,9.167)
(7.900,9.267)
(7.300,9.733)
(6.750,8.000)
(7.400,9.667)
(6.332,8.667)
(7.300,8.112)
(7.050,6.833)
(8.050,8.890)
(7.068,7.723)
(7.700,9.000)
(6.868,6.945)
(7.700,6.833)
(8.360,9.223)
(8.000,9.458)
};
\addplot[domain=5.4:8.8, samples=2, thick] {3.5251 + 0.5941*x};
\legend{Groups, OLS fit}
\end{axis}
\end{tikzpicture}
\caption{Scatter plot of the team-health index versus the average project grade. The fitted line corresponds to the simple linear regression model $\widehat{y} = 3.525 + 0.594x$.}
\label{fig:scatter}
\end{figure}

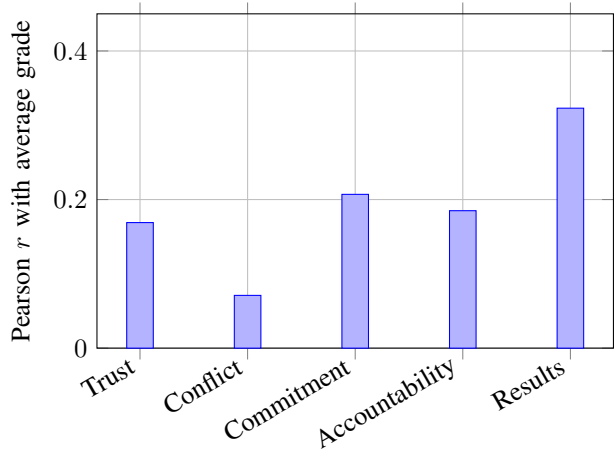
\begin{figure}[!t]
\centering
\begin{tikzpicture}
\begin{axis}[
    ybar,
    width=0.95\linewidth,
    height=6cm,
    ymin=0,
    ymax=0.45,
    grid=both,
    bar width=10pt,
    symbolic x coords={Trust,Conflict,Commitment,Accountability,Results},
    xtick=data,
    x tick label style={rotate=30,anchor=east},
    ylabel={Pearson $r$ with average grade},
]
\addplot coordinates {
    (Trust,0.169)
    (Conflict,0.071)
    (Commitment,0.207)
    (Accountability,0.185)
    (Results,0.323)
};
\end{axis}
\end{tikzpicture}
\caption{Pearson correlations between each team-dynamics dimension and the average project grade.}
\label{fig:bars}
\end{figure}

Overall, the evidence suggests a positive but limited relationship between team health and performance (Fig. \ref{fig:scatter} and \ref{fig:bars}). The results are strongest when the outcome is the teamwork competency, which is consistent with the idea that these items capture how the group works internally rather than the technical quality of the artifact alone.

\section{Discussion}

The results provide empirical evidence on the relationship between team dynamics and performance in a Capstone context, offering both confirmation and refinement of existing theoretical models.

\subsection{Interpretation of Results}

The \textit{Inattention to Results} dimension shows the strongest association with performance and is the only variable reaching statistical significance. This suggests that alignment around goals and outcomes may be more directly related to performance, although the observed effect size is modest.

In contrast, \textit{Absence of Trust}, which is central in Lencioni’s model, did not show a strong or statistically significant relationship with performance in this dataset. This indicates that the role of trust may be more complex or context-dependent in project-based learning environments.

The \textit{Lack of Commitment} dimension showed a moderate but non-significant association with performance, suggesting that clarity of direction and alignment around decisions may contribute to project outcomes, although the effect is limited.

The remaining dimensions—\textit{Fear of Conflict} and \textit{Avoidance of Accountability} presented weak associations with performance. One possible explanation is that these aspects may have more complex or indirect relationships with team effectiveness. For example, some degree of conflict may be beneficial when it reflects constructive debate rather than interpersonal tension.

\subsection{Alignment with Lencioni’s Model}

The findings provide only partial support for Lencioni’s framework. While the model emphasizes trust as the foundation of effective teamwork, the results of this study do not show a strong or statistically significant association between trust and performance.

Instead, the dimension most directly related to performance is \textit{Inattention to Results}, which aligns with the idea that outcome orientation plays a critical role in project success. This suggests that, in the context of Capstone projects, performance may be more strongly driven by goal alignment and focus on deliverables than by foundational interpersonal dynamics.

Furthermore, the expected cascading structure of the model—where dysfunctions influence one another—was not directly tested in this study and therefore cannot be empirically confirmed in this study. This indicates that the relationships among the dysfunctions may be more independent or context-dependent than originally proposed.

Overall, these findings suggest that Lencioni’s model may not fully explain performance in project-based learning environments, and that its applicability may depend on contextual factors such as evaluation criteria and project structure.

\subsection{Educational Implications}

From an educational perspective, these findings have important implications for the design and management of Capstone programs. First, the strong association between the \textit{Inattention to Results} dimension and performance suggests that instructors should emphasize goal alignment, clarity of expected outcomes, and shared responsibility for deliverables. Structured milestone tracking and explicit performance criteria may help reinforce this focus.

Second, although trust did not show a strong direct relationship with performance, it may still play an indirect role in enabling effective collaboration. Therefore, activities that support team cohesion and communication should not be neglected, but they may not be sufficient on their own to ensure high performance.

Third, the relatively weak associations observed for several dysfunctions suggest that technical and contextual factors remain critical. As a result, team dynamics should be considered as one component of a broader system influencing project success, rather than as the sole determinant.

\subsection{Limitations}

This study has several limitations that should be considered when interpreting the results. First, the data are based on self-reported perceptions, which may be subject to bias. Second, the relatively high concentration of scores in the upper range may limit the variability needed to detect stronger relationships. Third, the analysis is based on a single institution and a specific program structure, which may limit generalizability.

Additionally, the use of correlation analysis does not allow for causal inference. While associations can be identified, it is not possible to determine whether improvements in team dynamics directly lead to better performance. Furthermore, although this study focuses solely on final grades as the outcome measure, it is important to note that group dynamics may have been influenced by formative feedback provided during the activity, which is not explicitly accounted for in the analysis.

\section{Conclusion and Future Work}

This paper investigates the relationship between team dysfunctions and project performance in a Capstone program using group-level survey data and multi-dimensional performance evaluations. We examine the extent to which Lencioni’s model explains variation in team outcomes.

Results show a positive but limited association between team dynamics and performance. Among the five dimensions, \textit{inattention to results} (i.e., results orientation) exhibits the strongest and only statistically significant relationship, while the remaining dimensions show weaker or non-significant effects.

These findings suggest that alignment around goals and outcomes may play a more direct role in performance than other aspects of team dynamics. They also indicate that the dimensions of Lencioni’s model do not contribute equally to performance in this context.

From an educational perspective, the results highlight the importance of promoting goal alignment, clarity of expectations, and shared responsibility for deliverables. At the same time, team dynamics should be considered alongside technical and contextual factors when evaluating student performance.

Future work will extend the analysis across cohorts, incorporate longitudinal measurements, and evaluate targeted pedagogical interventions.

\bibliographystyle{ieeetr}
\bibliography{ref_local}

@book{lencioni2002five,
  title     = {The Five Dysfunctions of a Team: A Leadership Fable},
  author    = {Lencioni, Patrick M.},
  year      = {2002},
  publisher = {Jossey-Bass},
  address   = {San Francisco},
  edition   = {1},
  isbn      = {978-0787960759},
  pages     = {256},
  language  = {English},
  note      = {20th Anniversary Edition}
}

@article{whalley_computing_2025,
	title = {Computing {Capstone} {Courses} as {Preparation} for {Practice}: {A} {Global} {Survey} of {Instructors}},
	volume = {16},
	issn = {2153-2184, 2153-2192},
	shorttitle = {Computing {Capstone} {Courses} as {Preparation} for {Practice}},
	url = {https://dl.acm.org/doi/10.1145/3715880},
	doi = {10.1145/3715880},
	language = {en},
	number = {1},
	urldate = {2025-06-17},
	journal = {ACM Inroads},
	author = {Whalley, Jacqueline and Imbulpitiya, Asanthika and Clear, Tony},
	month = mar,
	year = {2025},
	pages = {26--39},
}

@INPROCEEDINGS{KhakurelPorras2020,
  author={Khakurel, Jayden and Porras, Jari},
  booktitle={2020 IEEE 32nd Conference on Software Engineering Education and Training (CSEE\&T)}, 
  title={The Effect of Real-World Capstone Project in an Acquisition of Soft Skills among Software Engineering Students}, 
  year={2020},
  volume={},
  number={},
  pages={1-9},
  doi={10.1109/CSEET49119.2020.9206201}
}

@article{jamaluddin_undergraduate_2013,
	title = {Undergraduate {Industrial} {Training} {Experience}: {A} {Win}-win {Situation} for {Students}, {Industry} and {Faculty}},
	volume = {102},
	copyright = {https://www.elsevier.com/tdm/userlicense/1.0/},
	issn = {18770428},
	shorttitle = {Undergraduate {Industrial} {Training} {Experience}},
	url = {https://linkinghub.elsevier.com/retrieve/pii/S1877042813043383},
	doi = {10.1016/j.sbspro.2013.10.783},
	language = {en},
	urldate = {2025-06-17},
	journal = {Procedia - Social and Behavioral Sciences},
	author = {Jamaluddin, Nordin and Ayob, Afida and Osman, Siti Aminah and Omar, Mohd Zaidi and Kofli, Norhisham Tan and Johar, Suhana},
	month = nov,
	year = {2013},
	pages = {648--653},
}

@article{nasir_software_2025,
	title = {Software engineering team project courses with industrial customers: {Students}’ insights on challenges and lessons learned},
	volume = {226},
	issn = {01641212},
	shorttitle = {Software engineering team project courses with industrial customers},
	url = {https://linkinghub.elsevier.com/retrieve/pii/S0164121225001098},
	doi = {10.1016/j.jss.2025.112441},
	language = {en},
	urldate = {2025-06-17},
	journal = {Journal of Systems and Software},
	author = {Nasir, Nayla and Usman, Muhammad and Börstler, Jürgen and Fogelström, Nina Dzamashvili},
	month = aug,
	year = {2025},
	pages = {112441},
}

@article{cecilia_bastarrica_impact_2023,
	title = {On the {Impact} of {Grading} on {Teamwork} {Quality} in a {Software} {Engineering} {Capstone} {Course}},
	volume = {11},
	issn = {2169-3536},
	url = {https://ieeexplore.ieee.org/document/10093802/},
	doi = {10.1109/ACCESS.2023.3265302},
	urldate = {2025-06-17},
	journal = {IEEE Access},
	author = {Cecilia Bastarrica, María and Gutierrez, Francisco J. and Marques, María and Perovich, Daniel},
	year = {2023},
	pages = {36492--36503},
}

@ARTICLE{FriessGoupee2020,
  author={Friess, Wilhelm A. and Goupee, Andrew J.},
  journal={IEEE Transactions on Education}, 
  title={Using Continuous Peer Evaluation in Team-Based Engineering Capstone Projects: A Case Study}, 
  year={2020},
  volume={63},
  number={2},
  pages={82-87},
  doi={10.1109/TE.2020.2970549}
}

@article{lopez_improving_2025,
	title = {Improving the {Assessment} of {Capstone} {Projects} in the {Bachelor}’s {Degree} in {Computer} {Engineering}},
	volume = {25},
	issn = {1946-6226},
	url = {https://dl.acm.org/doi/10.1145/3722230},
	doi = {10.1145/3722230},
	language = {en},
	number = {2},
	urldate = {2025-06-17},
	journal = {ACM Transactions on Computing Education},
	author = {Lopez, Angeles and Castaño, M. Asunción and Ibáñez, M. Victoria and Sanz, Ismael and Museros, Lledó and Grangel, Reyes},
	month = jun,
	year = {2025},
	pages = {1--28},
}

@inproceedings{garcia_application_2024,
	address = {Portland OR USA},
	title = {Application of {Collaborative} {Learning} {Paradigms} within {Software} {Engineering} {Education}: {A} {Systematic} {Mapping} {Study}},
	isbn = {979-8-4007-0423-9},
	shorttitle = {Application of {Collaborative} {Learning} {Paradigms} within {Software} {Engineering} {Education}},
	url = {https://dl.acm.org/doi/10.1145/3626252.3630780},
	doi = {10.1145/3626252.3630780},
	language = {en},
	urldate = {2025-06-17},
	booktitle = {Proceedings of the 55th {ACM} {Technical} {Symposium} on {Computer} {Science} {Education} {V}. 1},
	publisher = {ACM},
	author = {Garcia, Rita and Treude, Christoph and Valentine, Andrew},
	month = mar,
	year = {2024},
	pages = {366--372},
}

@inproceedings{grasl_exposing_2023,
	title = {Exposing {Software} {Engineering} {Students} to {Stressful} {Projects}: {Does} {Diversity} {Matter}?},
	shorttitle = {Exposing {Software} {Engineering} {Students} to {Stressful} {Projects}},
	url = {https://ieeexplore.ieee.org/document/10172779/},
	doi = {10.1109/ICSE-SEET58685.2023.00026},
	urldate = {2025-06-17},
	booktitle = {2023 {IEEE}/{ACM} 45th {International} {Conference} on {Software} {Engineering}: {Software} {Engineering} {Education} and {Training} ({ICSE}-{SEET})},
	author = {Graßl, Isabella and Fraser, Gordon and Trieflinger, Stefan and Kuhrmann, Marco},
	month = may,
	year = {2023},
	note = {ISSN: 2832-7578},
	pages = {210--222},
}

@inproceedings{porquet-lupine_evaluating_2023,
	address = {Toronto ON Canada},
	title = {Evaluating {Group} {Work} in (too) {Large} {CS} {Classes} with (too) {Few} {Resources}: {An} {Experience} {Report}},
	isbn = {978-1-4503-9431-4},
	shorttitle = {Evaluating {Group} {Work} in (too) {Large} {CS} {Classes} with (too) {Few} {Resources}},
	url = {https://dl.acm.org/doi/10.1145/3545945.3569788},
	doi = {10.1145/3545945.3569788},
	language = {en},
	urldate = {2025-06-16},
	booktitle = {Proceedings of the 54th {ACM} {Technical} {Symposium} on {Computer} {Science} {Education} {V}. 1},
	publisher = {ACM},
	author = {Porquet-Lupine, Joël and Brigham, Madison},
	month = mar,
	year = {2023},
	pages = {4--10},
}

@article{paretti13,
    author = {Paretti, Marie and Pembridge, J.J. and Brozina, S.C. and Lutz, Ben and Phanthanousy, J.N.},
    year = {2013},
    month = {01},
    pages = {},
    title = {Mentoring team conflicts in capstone design: Problems and solutions},
    journal = {ASEE Annual Conference and Exposition, Conference Proceedings}
}

@mastersthesis{edwards2012team,
  author       = {Edwards, Calvin W.},
  title        = {A Case Study of the Adaptation of a Team Building Model Using Action Learning},
  school       = {University of Pennsylvania},
  year         = {2012},
  address      = {Philadelphia, PA, USA},
  type         = {Master's Thesis},
  note         = {Master of Science in Organizational Dynamics}
}

@INPROCEEDINGS{6462328,
  author={Pohopien, Laura and Hogan, Gina and Bayne, Stephan and Temple, James and Fiero, Diane and Devlin, Allison and Patrick, John and Sexton, Nate and Brooks, Jalin and Stein, Penny and Arty, Anthony and Luechtefeld, Ray},
  booktitle={2012 Frontiers in Education Conference Proceedings}, 
  title={Trust in engineering teams and groups and virtual facilitation methods}, 
  year={2012},
  volume={},
  number={},
  pages={1-6},
  doi={10.1109/FIE.2012.6462328}}

@inproceedings{soares2018,
	title = {Continuous integrated team learning},
	booktitle = {Proceedings of the {PAEE}/{ALE}’2018, 10th international symposium on project approaches in engineering education ({PAEE}) and 15th active learning in engineering education workshop ({ALE})},
	author = {Soares, L. P. and Ferrão, R. C.},
	year = {2018},
	pages = {202--209},
}

@inproceedings{ferrao2025,
  author    = {Ferr{\~a}o, Rafael Corsi and Soares, Luciano},
  title     = {Enhancing Capstone Program Workflow: A Case Study on a Platform for Managing Academic-Industry Projects},
  booktitle = {Proceedings of {PAEE}-{ALE}},
  address   = {Porto, Portugal},
  year      = {2025},
  doi       = {10.5281/zenodo.15850131}
}

\end{document}